\documentstyle{aipproc}
\input epsf
\input psfig

\begin{document}
\hfill INP Cracow preprint 1830/PH

\title{Mesons in non-local chiral quark models\thanks{%
Research supported by the Polish State Committee for Scientific Research,
grant 2P03B-080-12, and by DSF and BMBF.}}

\author{Wojciech Broniowski}

\address{H. Niewodnicza\'nski Institute of Nuclear Physics,
 PL-31-342 Krak\'ow, Poland\\
{\rm broniows@solaris.ifj.edu.pl}}
\maketitle

\begin{abstract}
After briefly reviewing chiral quark models with non-local regulators and
listing their advantages over the conventional Nambu-Jona-Lasionio-like
models, we study vector meson correlators in both types of
approaches. Since
effective chiral quark models are valid in the Eulidean domain only, with $%
-0.5{\rm GeV}^2\leq q^2 \leq 0$, in our study the meson correlators are
not described directly, but with help of a method based on dispersion
relations for the physical correlators. A set of sum rules is derived, which
allows for a comparison of model predictions to data. We find that both the
local and non-local models fail to satisfy the sum rules,
unless very low values of the constituent quark
mass parameter are used.
We also show that the two Weinberg sum rules hold in the
non-local model.
\end{abstract}


This research has been done in collaboration with Maxim
Polyakov from Bochum.

\section{Why non-local regulators?}

First, let me recall the reasons why we wish to consider chiral quark models
with non-local regulators (see also the contribution by Bojan Golli and
Georges Ripka to these proceedings):

\begin{enumerate}
\item  Non-locality arises naturally in several approaches to low-energy
quark dynamics, such as the instanton-liquid model \cite
{Diakonov86,instant1:rev,instant2:rev} or Schwinger-Dyson resummations \cite
{Roberts92}. For the discussions of various ``derivations'' and applications
of non-local quark models see, {\em e.g.},\cite
{Ripka97,Cahill87,Holdom89,Ball90,Krewald92,Ripka93,BowlerB,PlantB,mitia:rev,Bijnens}%
. Hence, we should cope with non-local regulators from the outset.

\item  Non-local interactions regularize the theory in such a way that the
anomalies are preserved \cite{Ripka93,Arrio} and charges are properly
quantized. With other methods, such as the proper-time regularization or the
quark-loop momentum cut-off \cite{mitia:rev,Bijnens,Goeke96,Reinhardt96} the
preservation of the anomalies can only be achieved if the (finite) anomalous
part of the action is left unregularized, and only the non-anomalous
(infinite) part is regularized. If both parts are regularized, anomalies are
violated badly \cite{krs,bbs}. We consider such a division artificial and find
it quite appealing that with non-local regulators both parts of the action
can be treated on equal footing.

\item  With non-local interactions the effective action is finite to all
orders in the loop expansion ($1/N_c$ expansion). In particular, meson loops
are finite and there is no more need to introduce extra cut-offs, as was
necessary in the case of local models \cite{Ripka96b,Tegen95,Temp}.
As the result, non-local models have more predictive power.

\item  As Bojan Golli, Georges Ripka and WB have shown \cite{nls}, stable
solitons exist in a chiral quark model with non-local interactions without
the extra constraint that forces the $\sigma$ and $\pi$ fields to lie on
the chiral circle. Such a constraint is external to the known derivations
of effective chiral quark models.

\item  The empirical values of the low-energy constants $g_8$ and $g_{27}$
of the effective weak chiral Lagrangian are better reproduced within the
non-local model \cite{Franz} compared to the conventional NJL model.
\end{enumerate}

In view of these improvements it becomes important to look at all other
applications of effective quark models and compare the predictions of
non-local and local versions.

\section{Dispersion-relation sum rules for meson correlators}

In this talk we will present results for the
vector-meson correlators.
\begin{figure}[bp]
\vspace{0mm} \epsfxsize = 11 cm \centerline{\epsfbox{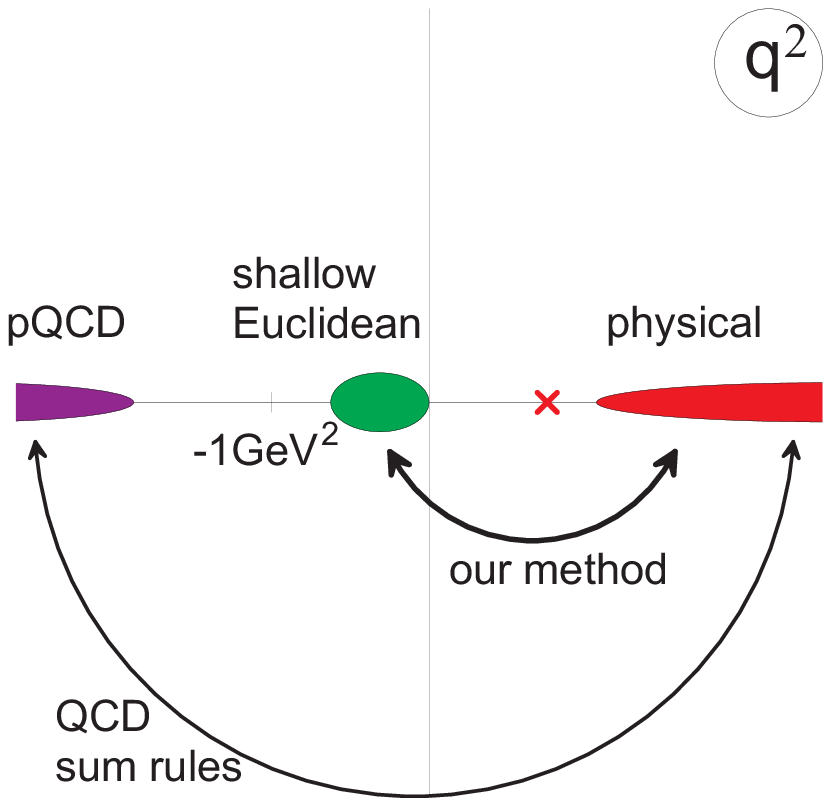}} %
\vspace{12mm} \label{analyt}
\caption{}
\label{fig:shallow}
\end{figure}
The basic object of our study is the meson correlation function, defined as
\begin{equation}
\Pi^{AB}(q^2)=\langle 0|i\int d^4x\,e^{iq\cdot x}{\rm T}\left\{
j^A(x),j^B(0)\right\} |0\rangle ,  \label{corr}
\end{equation}
where ${\rm T}$ denotes the time-ordered product, and $j^A(x)$ describes a
color-singlet quark bilinear with appropriate Lorentz and flavor matrix,
{\em e.g.} in the $\rho$-meson channel we have $j^a_\mu (x)=\frac{1}{2} \bar
\psi (x)\gamma_\mu \tau^a \psi (x)$. The tensor structure can be taken away,
with $\Pi ^{AB}(q^2)=t^{AB}\Pi (q^2)$, where $t^{AB}$ is a tensor in the
Lorentz and isospin indices, and $\Pi (q^2)$ is a scalar function. The
powerful feature of $\Pi (q^2)$ is its analyticity in the $q^2$ variable,
which will be used shortly. In Figure 1 we display various regions in the
complex $q^2$ plane: the positive real axis is the physical region, with
poles and cuts corresponding to physical states in the particular channel.
In that physical region we have (for certain channels) direct experimental
information. Far to the left, at the end of the negative real axis, is the
deep-Euclidean region, where perturbative QCD and its operator product
expansion can be applied. There is yet another region, close to $0$ on the
negative real axis: the {\em shallow-Euclidean} region. This is the
playground of the effective chiral models. Indeed, the most successful
attempt to derive such a model from QCD, namely, the instanton-liquid model,
has a natural limitation to that region of momenta \cite
{instant1:rev,instant2:rev}.
More to the point, 
we believe that all effective chiral quark models should be used 
{\em in and only in} the shallow Euclidean domain. There have been many
attempts, however, to apply such models directly to the physical region. In
our opinion these are doomed to fail because of the lack of confinement,
which is a key player in the physical region. With unconfined quarks, the
unphysical $q\bar q$ continuum obstructs any model calculation for $%
q^2>(2M)^2$, where $M\sim 300-400{\rm MeV}$ is the ``constituent'' quark
mass. For these reasons of principles we always remain with the model at low
negative $q^2$ \cite{Boch,tensor}, and compare to the data via the
dispersion relation which holds for the {\em physical} correlator. In other
words, we bring the physical data to the shallow Euclidean region via the
dispersion relation. This is similar in spirit to the QCD sum rule approach,
where one compares the physical spectrum to the deep Euclidean region via
(Borelized) dispersion relation (see Figure 1).

The correlators considered here satisfy the twice-subtracted dispersion
relation 
\begin{equation}
\Pi (Q^2)=c_0+c_1Q^2+\frac{Q^4}\pi \int_0^\infty ds\frac{{\rm Im}\Pi (s)}{%
s^2(s+Q^2)},  \label{disprel}
\end{equation}
where $Q^2=-q^2$, and $c_i$ are subtraction constants. Relation (\ref
{disprel}) holds for the {\em physical} correlators, and does not
in general hold for the correlators evaluated in models \cite{analyt}. For
some channels (vector channels) ${\rm Im}\Pi (s)$ is obtained directly from
experiment, in other channels we have indirect information only, {\em e.g.}
from QCD\ sum rules. Let us take the $\rho $-meson channel, where
$j_a^\mu (x)=\frac{1}{2} \bar \psi (x)\gamma ^\mu \tau_a\psi (x)$ 
and $\Pi _{ab}^{\mu \nu }(Q^2)=\delta _{ab}(Q^\mu Q^\nu /Q^2-g^{\mu \nu
})\Pi _\rho (Q^2)$. The spectral strength in related to the ratio 
\begin{equation}
{\rm Im}\Pi _\rho ^{{\rm phen}}(s)=\frac s{6\pi }\frac{\sigma
(e^{+}e^{-}\rightarrow n\pi )}{\sigma (e^{+}e^{-}\rightarrow \mu ^{+}\mu
^{-})},\quad n=2,4,6,...  \label{ratio}
\end{equation}
known very accurately from experiment. The spectral strength peaks at the
position of the $\rho $-meson pole, and at large $s$ assumes the
perturbative-QCD value. For our task it is more convenient to use
the simple pole+continuum parameterization of ${\rm Im}\Pi _\rho (s)$, such
as used {\em e.g.} in QCD\ sum rules. In a given channel the fit has the
form 
\begin{equation}
{\rm Im}\Pi ^{{\rm phen}}(s)=\frac{\pi s^2}{g^2}\delta \left( s-m^2\right)
+a\,s\,\theta (s-s_0),  \label{fitrho}
\end{equation}
where $a$ is known from perturbative QCD, and $g$, $m$ and $s_0$ are chosen
such that the experimental data are reproduced. In the $\rho$-channel we
have $a=\frac 1{8\pi }(1+\alpha _s/\pi +...)$, $m=0.77{\rm
GeV}$, $g^2/(4\pi )=2.36$, and $s_0=1.5{\rm GeV}^2$ \cite{QCD:sum:1}.
Since all our calculations will be done in the leading-$N_c$
level, we drop the $\alpha _s$ correction in $a$.

With the parametrization (\ref{fitrho}) we readily obtain from (\ref{disprel}%
)
\begin{equation}
\Pi ^{{\rm phen}}(Q^2)=c_0+c_1Q^2+\frac{Q^4}{g^2(m^2+Q^2)}+\frac a\pi
Q^2\log \left( 1+\frac{Q^2}{s_0}\right) .  \label{rhophen}
\end{equation}
On the other hand, $\Pi (Q^2)$ can be calculated directly in chiral quark
models in the shallow Euclidean space. We denote this model correlation $\Pi
^{{\rm mod}}(Q^2),$ as want to compare it somehow to $\Pi ^{{\rm phen}}(Q^2)$%
. One possibility is to Fourier-transform to coordinate space \cite
{Shuryak93,Shuryak93b,Arriola95b,JamArr}. Here we apply a simpler method,
which relies on just Taylor-expanding $\Pi ^{{\rm phen}}(Q^2)$ and $\Pi ^{%
{\rm mod}}(Q^2)$ in the $Q^2$ variable. For the phenomenological correlator
we get from Eq. (\ref{rhophen}) 
\begin{equation}
\Pi ^{{\rm phen}}(Q^2)=\sum_{k=1}^\infty (-)^{k+1}b_k^{{\rm phen}%
}=c_0+c_1Q^2+Q^2\sum_{k=1}^\infty (-)^{k+1}\left[ \left( \frac{Q^2}{m^2}%
\right) ^k+\frac a{k\pi }\left( \frac{Q^2}{s_0}\right) ^k\right] ,
\label{rhophexp}
\end{equation}
whereas for the model correlator we can write 
\begin{equation}
\Pi ^{{\rm mod}}(Q^2)=\sum_{k=1}^\infty (-)^{k+1}b_k^{{\rm mod}},
\label{rhomodexp}
\end{equation}
with the expansion 
coefficients $b_k^{{\rm mod}}$ yet to be determined. We can now
compare the coefficients $b_k^{{\rm phen}}$ and $b_k^{{\rm mod}}$, and form
a set of ``sum rules''. With two subtractions in (\ref{disprel}) we can
start at $k=2$: $b_k^{{\rm phen}}=b_k^{{\rm mod}}$, $k\geq 2$. With the
explicit form (\ref{rhophexp}) this gives 
\begin{equation}
b_2^{{\rm mod}}=\frac 1{g^2m^2}+\frac a{\pi s_0},\quad b_3^{{\rm mod}}=\frac
1{g^2m^4}+\frac a{2\pi s_0^2},\quad ...  \label{cqsr}
\end{equation}
Sum rules for higher values of $k$ are sensitive to the details of the
phenomenological spectrum, hence are not going to be of much help. 

In some
channels the coupling constant $g$ is not well known. We can then eliminate
it from Eqs. (\ref{cqsr}) to obtain 
\begin{equation}
m^2=\frac{b_2^{{\rm mod}}-\frac a{\pi s_0}}{b_3^{{\rm mod}}-\frac a{2\pi
s_0^2}},\quad m^2=\frac{b_3^{{\rm mod}}-\frac a{2\pi s_0^2}}{b_4^{{\rm mod}%
}-\frac a{3\pi s_0^3}},\quad ...  \label{cqsr2}
\end{equation}
Sum rules (\ref{cqsr}) or (\ref{cqsr2}) can be used to verify model
predictions for meson correlators.

\section{Results for the local model}

The model evaluation of meson correlators is well known. At the leading-$N_c$
level one has

\centerline{\psfig
{%
figure=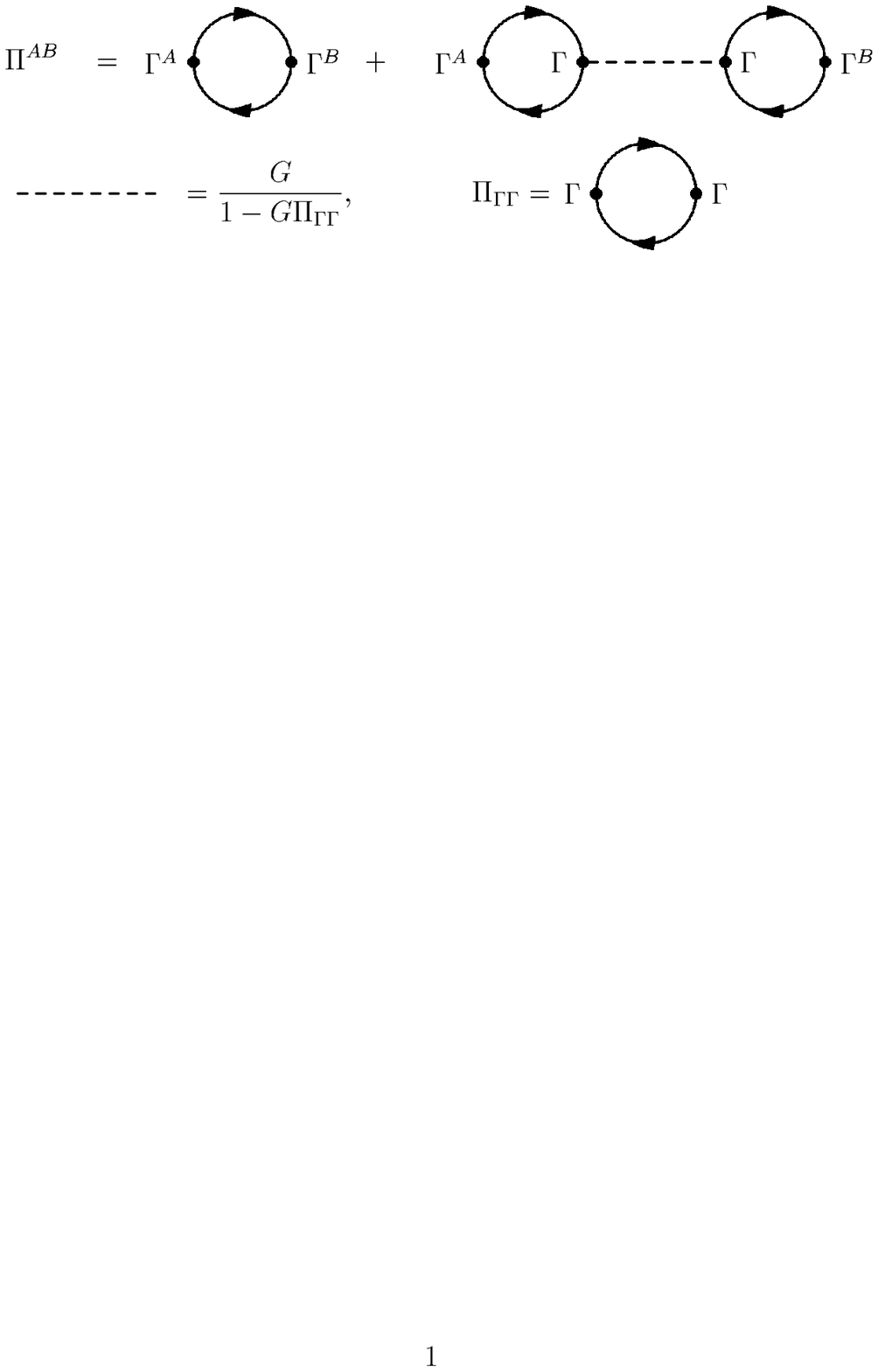,height=40mm,bbllx=120bp,bblly=580bp,bburx=490bp,bbury=700bp,clip=%
}} 
\noindent The diagrams with the dashed line occur only if the coupling
constant $G$ is non-zero in a given channel. This is the case of the of the $%
\sigma $ and $\pi $ channels. In the vector channels they may or may not be
present, depending on whether we allow for explicit vector interactions
between the quarks.

We first consider the $\rho$-channel in the local NJL model with the
proper-time regularization \cite{mitia:rev,Bijnens,Goeke96,Reinhardt96}.
After some simple algebra we get 
\begin{equation}
b_2^{{\rm mod}}=\frac 1{8\pi ^2}e^{-M^2/\Lambda ^2}\frac 1{5M^2},\quad b_3^{%
{\rm mod}}=\frac 1{8\pi ^2}e^{-M^2/\Lambda ^2}\frac{3(M^2+\Lambda ^2)}{%
140M^4\Lambda ^2},\quad ...  \label{PT}
\end{equation}
where $M$ is the 
constituent quark mass generated by the spontaneous 
breaking of the chiral symmetry, and $\Lambda $ is the proper-time cut-off,
adjusted such that the pion decay constant has its experimental value, $%
F_\pi =93{\rm MeV}$. Here we work in the strict chiral limit, with the
current quark mass set to zero. The results for sum rules (\ref{cqsr2}) are
shown in Table \ref{rhoNJL0}. We can see that the ratios of phenomenological
to model coefficients $b_2$ and $b_3$ are larger than $1$, and increase
rather rapidly with increasing $M$. Thus the sum rules (\ref{cqsr}) favor
lower values of $M$. However, even for such low values as $M=250{\rm MeV}$
the ratios $b_2^{{\rm phen}}/b_2^{{\rm mod}}$ and $b_3^{{\rm phen}}/b_3^{%
{\rm mod}}$ are still significantly above $1$. We conclude that the model is
far from satisfying the sum rules (\ref{cqsr}).

\begin{table}[h]
\caption{$\rho $ channel sum rules in the NJL model with the proper-time
regulator ($m=0$, $F_\pi=93{\rm MeV}$).}
\label{rhoNJL0}%
\begin{tabular}{cddd}
$M$ [GeV] & $\Lambda$ [GeV] & $b_2^{\rm phen}/b_2^{\rm mod}$ &
$b_3^{\rm phen}/b_3^{\rm mod}$  \\
\tableline
0.25 & 0.79 & 1.8 & 1.4 \\
0.3 & 0.69 & 2.8 & 3.0 \\
0.35 & 0.65 & 4.2 & 5.7 \\
0.4 & 0.64 & 6.1 & 9.9 \\
\end{tabular}
\end{table}

Next, we repeat our calculation for the variant of the model where vector
interactions are included \cite{Bijnens,Klimt,Vogl} in the Lagrangian: $-\frac
12G_\rho \left( (\bar \psi \gamma _\mu \tau ^a\psi )^2+(\bar \psi \gamma
_\mu \gamma _5\tau ^a\psi )^2\right) $. In that model the formulas for the
axial coupling constant of the quark, $g_A^Q$, and for $F_\pi $ read 
\begin{equation}
g_A^Q=\left( 1+G_\rho \frac{N_cM^2}{\pi ^2}\Gamma (0,M^2/\Lambda ^2)\right)
^{-1},\quad F_\pi ^2=g_A^Q\frac{N_cM^2}{4\pi ^2}\Gamma (0,M^2/\Lambda ^2),
\label{gafpi}
\end{equation}
with $\Gamma (0,x)=\int_x^\infty dt\,e^{-t}/t$. The results are shown in
Table \ref{rhoNJLG}. We can see that the ratio $b_2^{{\rm phen}}/b_2^{{\rm %
mod}}$ decreases as $G_\rho $ increases. However, uncomfortably large values
of $G_\rho $ are needed in order to satisfy the sum rule,{\em \ i.e.} to
make $b_2^{{\rm phen}}/b_2^{{\rm mod}}\sim 1$. The conclusion is that at
moderate values of $M$ the model needs very large values of $G_\rho $ to
describe properly the vector channel.

\begin{centering}
\begin{table}[h]
\caption{Same as Table \ref{rhoNJL0} with vector interactions included.}
\label{rhoNJLG}%
\begin{tabular}{r|dddd|dddd}
\multicolumn{1}{c|}{} & \multicolumn{4}{c|}{$M=0.3{\rm GeV}$}  &
   \multicolumn{4}{c}{$M=0.35{\rm GeV}$} \\
\tableline
$G_\rho$ [GeV$^{-2}$] &
$\Lambda$ [GeV] & $\frac{b_2^{\rm phen}}{b_2^{\rm mod}}$ &
$\frac{b_3^{\rm phen}}{b_3^{\rm mod}}$ & $g_A^Q$ &
$\Lambda$ [GeV] & $\frac{b_2^{\rm phen}}{b_2^{\rm mod}}$ &
$\frac{b_3^{\rm phen}}{b_3^{\rm mod}}$ & $g_A^Q$ \\
\tableline
0  & 0.69 & 1.8 & 3.0 & 1    & 0.65 & 4.2 & 5.7 & 1    \\
4  & 0.78 & 2.2 & 2.1 & 0.86 & 0.72 & 3.4 & 4.0 & 0.86 \\
8  & 0.91 & 1.6 & 1.4 & 0.72 & 0.81 & 2.5 & 2.8 & 0.72 \\
12 & 1.14 & 1.0 & 1.0 & 0.58 & 0.97 & 1.7 & 1.9 & 0.58 \\
\end{tabular}
\end{table}
\end{centering}

\section{Results for the non-local model}

The {\em non-local chiral quark model} differs from the local versions in
the fact that the interaction vertex carries momentum-dependent factors $%
r(p_i)$:

\centerline{\psfig
{%
figure=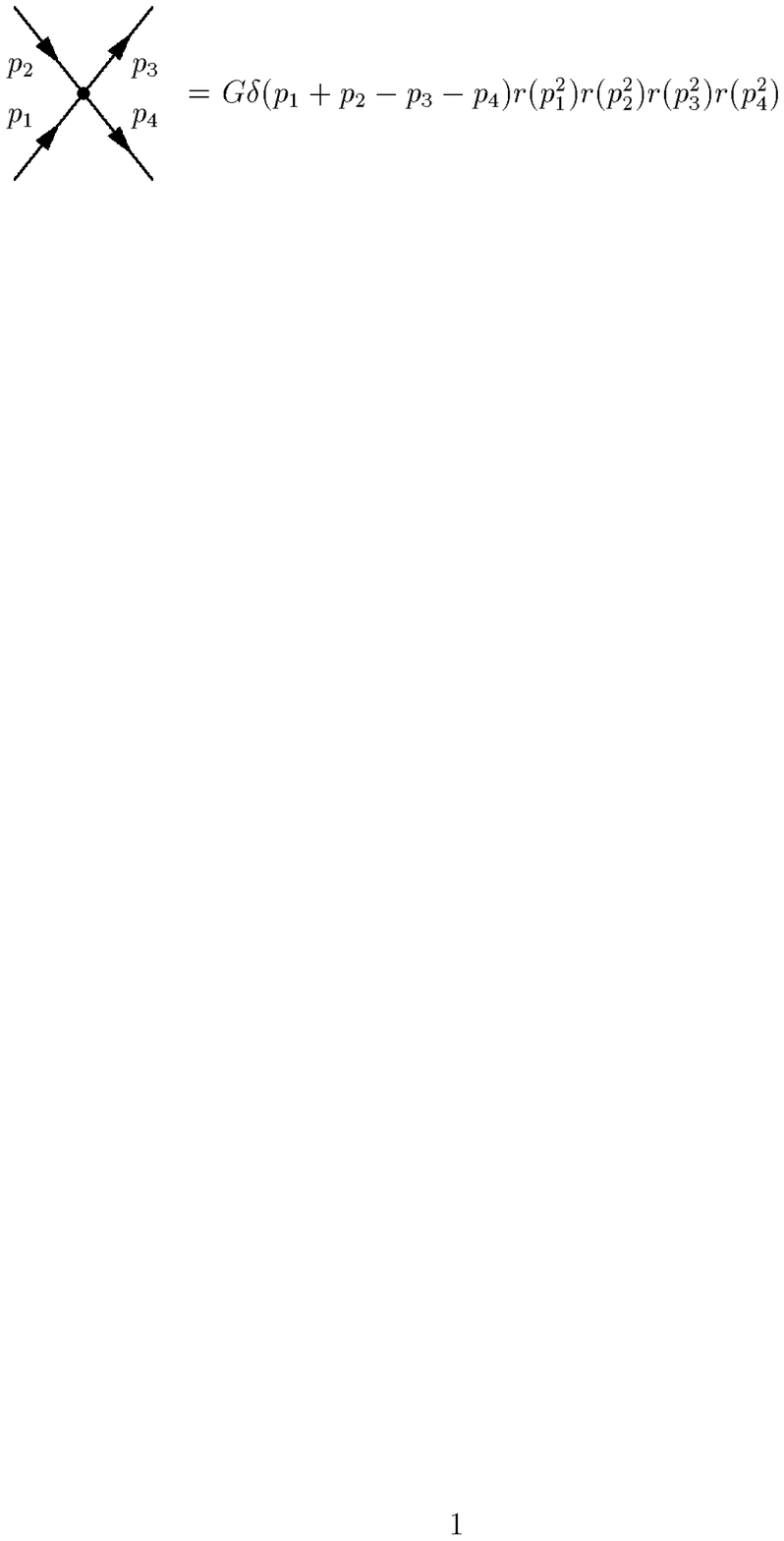,height=25mm,bbllx=120bp,bblly=645bp,bburx=430bp,bbury=717bp,clip=%
}} 
\noindent For some more details see the contribution of Bojan Golli and
Georges Ripka. We use here the following form of the regulator, $%
r(p^2)=1/(1+p^2/\Lambda ^2)^2$, which is simpler that the instanton-model
expression but is known to reproduce well its basic predictions. We use the
notation $M_k=M r(k^2)^2$, $D_k=k^2+$ $M_k^2$. In the vacuum sector one
finds that 
\begin{eqnarray}
\frac 1G &=&4N_cN_f\int \frac{d_4k}{(2\pi )^4}\frac{r(k^2)^4}{D_k},\quad
\langle \bar uu\rangle =\langle \bar dd\rangle =-4N_c\int \frac{d_4k}{(2\pi
)^4}\frac{M_k}{D_k},  \label{vacfits} \\
\langle \frac{\alpha _s}{8\pi }G_{\mu \nu }^aG_a^{\mu \nu }\rangle
&=&4N_c\int \frac{d_4k}{(2\pi )^4}\frac{M_k^2}{D_k},\quad F_\pi ^2=4N_c\int 
\frac{d_4k}{(2\pi )^4}\frac{M_k^2-k^2M_kM_k^{\prime }+k^4M_k^{\prime 2}}{%
D_k^2},  \nonumber
\end{eqnarray}
where the first equation is the stationary point condition, expressing the
quark coupling constant via the parameter $M$, the second equation is the
quark condensate in the chiral limit, the third is the gluon condensate in
the chiral limit, \cite{Weiss}, and the last one gives the pion decay
constant \cite{BowlerB,PlantB} in the chiral limit. We use the notation $%
M_k^{\prime }=dM_k/dk^2$.

There is a complication associated with non-local interactions, namely the
Noether currents pick up extra contributions. Furthermore, the transverse
parts of these currents are not uniquely determined and their choice
constitutes a part of the model. Here we use the so-called straight-line $P$%
-exponent prescription for gauging the model. For a discussion of this and
related issues see Refs. \cite{BowlerB,PlantB,Bled99}. We then find that

\centerline{\psfig
{%
figure=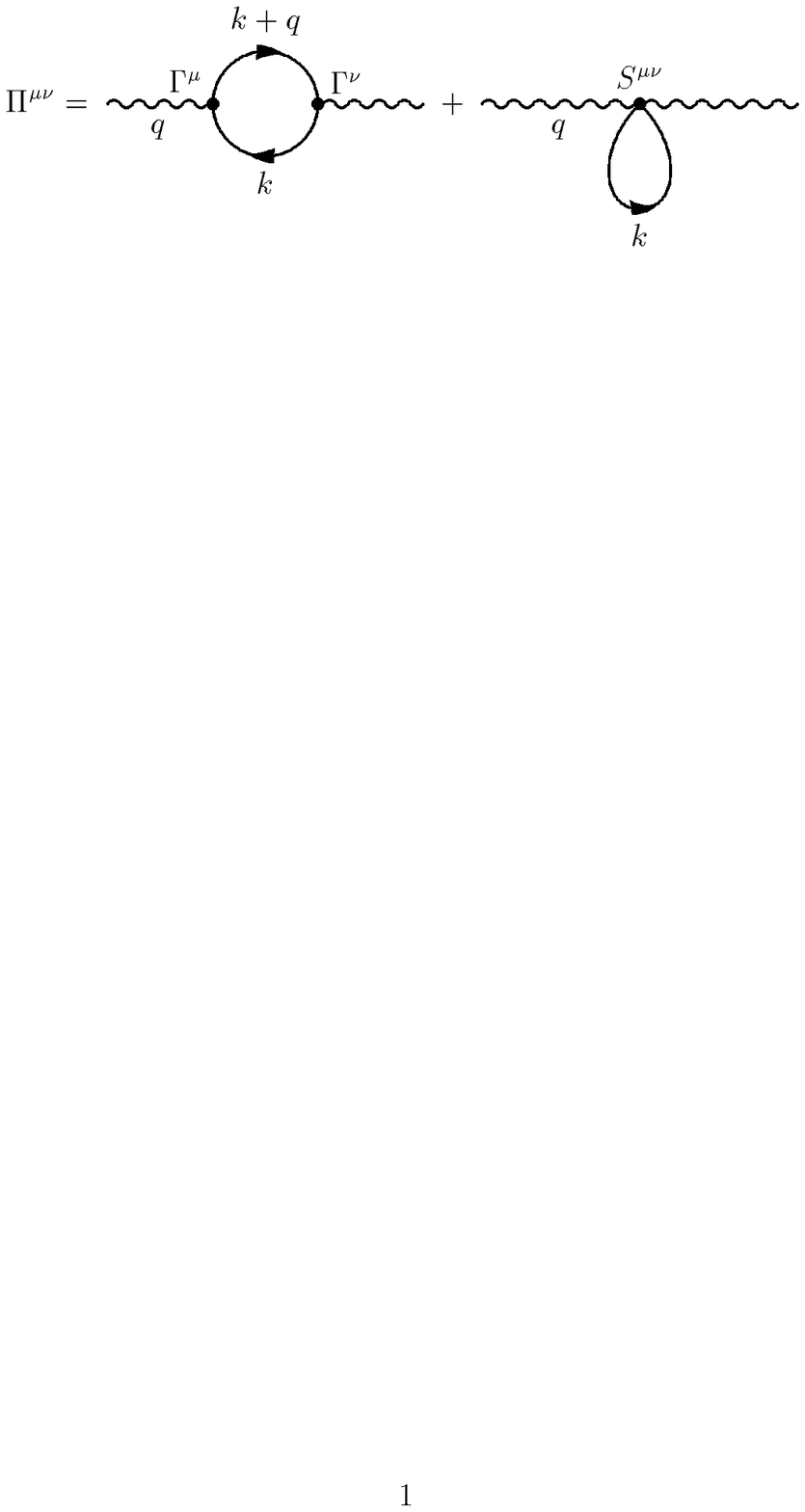,height=35mm,bbllx=144bp,bblly=612bp,bburx=460bp,bbury=710bp,clip=%
}} where 
\begin{equation}
\Gamma ^\mu =\gamma ^\mu -\int_0^1d\alpha \,\frac{dM(k+\alpha q)}{dk_\mu }%
,\quad S^{\mu \nu }=-\int_0^1d\alpha \int_0^1d\beta \,\frac{d^2M(k+\alpha
q-\beta q)}{dk_\mu dk_\nu }.  \label{gamsig}
\end{equation}
It is simple to check that the current is conserved, $q_\mu \Pi ^{\mu \nu
}=0 $. 
\begin{table}[h]
\caption{Same as Table \ref{rhoNJL0} for the non-local model.}
\label{rhonl}%
\begin{tabular}{dddd}
$M$ [GeV] & $\Lambda$ [GeV] & $b_2^{\rm phen}/b_2^{\rm mod}$ &
$b_3^{\rm phen}/b_3^{\rm mod}$  \\
\tableline
0.25 & 1.35 & 1.5 & 1.4 \\
0.3 & 1.0 & 1.9 & 2.9 \\
0.35 & 0.83 & 2.2 & 3.9 \\
0.4 & 0.72 & 2.3 & 3.8 \\
\end{tabular}
\end{table}
Coming back to the sum rules (\ref{cqsr}), we note by comparing Tables \ref
{rhoNJL0} and \ref{rhonl} that the results are a bit better in the non-local
model. Especially at larger values of $M$, around 350MeV, we gain about a
factor of 2. The discrepancy leaves room for such effects as the
vector-channel interactions and $1/N_c$ corrections, which should be the
object of a further study.

The effects of non-localities in currents, as depicted in the figure for $%
\Pi ^{\mu \nu }$, bring about 15 \% to the results. More precisely, the
calculation with $\gamma ^\mu $ instead of $\Gamma ^\mu $ in the vertices
(and without the sea-gull term $S^{\mu \nu }$) yields $b_2^{{\rm mod}}$ and $%
b_3^{{\rm mod}}$ roughly 15\% lower.

\section{Weinberg sum rules}

Now we turn to a more formal aspect of our study. An appealing feature of
non-local regulators is that now both Weinberg sum rules hold. The famous
sum rules (I) and (II) are: 
\begin{eqnarray}
\frac 1\pi \int_0^\infty \frac{ds}s\left[ {\rm Im}\Pi _\rho (s)-{\rm Im}\Pi
_{A_1}(s)\right] &=&F_\pi ^2,  \eqnum{I} \\
\frac 1\pi \int_0^\infty ds\left[ {\rm Im}\Pi _\rho (s)-{\rm Im}\Pi
_{A_1}(s)\right] &=&-m\langle \bar uu+\bar dd\rangle .  \eqnum{II}
\label{wsrdwa}
\end{eqnarray}
Whereas (I) holds in all variants of the NJL model, in local models (II)
picks up $M$ instead of $m$ on the right-hand side, thus is violated badly.
To prove the sum rules in the non-local model we consider the dispersion
relation 
\begin{equation}
\Pi _\rho (Q^2)-\Pi _{A_1}(Q^2)=\frac 1\pi \int_0^\infty \frac{ds}{s+Q^2}%
\left[ {\rm Im}\Pi _\rho (s)-{\rm Im}\Pi _{A_1}(s)\right] .  \label{diWein}
\end{equation}
No subtractions are necessary. We set $Q^2=0$. An explicit evaluation gives 
\begin{equation}
\Pi _\rho (0)-\Pi _{A_1}(0)=4N_c\int \frac{d_4k}{(2\pi )^4}\frac{%
M_k^2-k^2M_kM_k^{\prime }+k^4M_k^{\prime 2}}{D_k^2},  \label{checkone}
\end{equation}
in which we recognize our formula for $F_\pi ^2$ (\ref{vacfits}), thus
verifying (I). To prove WSR II we multiply both sides of (\ref{diWein}) by $%
Q^2$ and take the limit of $Q^2\rightarrow \infty $. We find, to the first
order in the current quark mass $m$, 
\begin{eqnarray}
&&\lim_{Q^2\rightarrow \infty }Q^2\left( \Pi _\rho (Q^2)-\Pi
_{A_1}(Q^2)\right) = \nonumber \\
&& \lim_{Q^2\rightarrow \infty }Q^2\int \frac{d_4k}{(2\pi )^4}\frac{%
4N_c[M_kM_{k+Q}+m(M_k+M_{k+Q})]}{D_kD_{k+Q}}=  \nonumber \\
&& 4N_cm\lim_{Q^2\rightarrow \infty }Q^2\int \frac{d_4k}{(2\pi )^4}\left( 
\frac{M_k}{D_kD_{k+Q}}+\frac{M_k}{D_kD_{k-Q}}\right) = \nonumber \\
&& 8mN_c\int \frac{d_4k}{(2\pi )^4}\frac{M_k}{D_k}=-2m\langle \bar uu+\bar
dd\rangle .  \label{W2der}
\end{eqnarray}
In passing from the second to the third line in the above derivation we have
used the fact that $M_p$ is strongly concentrated around $p=0$, thus we
could drop the term with $M_kM_{k+Q}$ at $Q^2\gg \Lambda ^2$. By a similar
argument in the third line we have replaced $Q^2/\left( (k\pm
Q)^2+M_{k+Q}^2\right) $ by $1$. Finally, in the last line we have recognized
our expression for the quark condensate (\ref{vacfits}). We note that
non-local contributions to the Noether currents are suppressed and do not
contribute to (\ref{W2der}). Similarly, rescattering diagrams as displayed
in the equation for $\Pi^{AB}$ can be dropped. This is because the vertex $%
\Gamma$ contains the regulator, hence the diagram is strongly suppressed at
large $Q^2$.

Clearly, the reason for the compliance with the second Weinberg sum rule is
the fact that the momentum-dependent mass of the quark becomes
asymptotically, in the deep-Euclidean region, just the current quark mass.
This is not the case of local models \cite{Bijnens,Dmi}, where the mass is
constant, and this is why these models violate Eq. (\ref{wsrdwa}).

\section{Final remarks}

To end this talk we describe shortly the results for other channels. In the $%
\omega$-meson channel the model results are, at the leading-$N_c$ level,
exactly the same as for the $\rho$-channel. In the pseudoscalar and scalar
channels we do not know the the corresponding parameter $g$ from experiment,
hence we consider sum rules (\ref{cqsr2}). In the pion channel the pion pole
entirely dominates the sum rules, {\em i.e.} the continuum contribution is
negligeable and we get $m_\pi^2 \simeq b_2^{{\rm mod}}/b_3^{{\rm mod}}
\simeq b_3^{{\rm mod}}/b_4^{{\rm mod}}...$ in both the local and non-local
variants of the model. In the $\sigma$-meson channel things are more
interesting. Whereas in the local model the sum rules (\ref{cqsr2}) simply
give the pole at twice the quark mass, $m_\sigma=2M$, in the non-local model
the predicted value of $m_\sigma$ ranges from 400MeV at $M=300{\rm MeV}$ to
470MeV at $M=450{\rm MeV}$ and is insensitive to the value of the threshold
parameter $s_0$.

One of us (WB) is grateful to Bojan Golli, Georges Ripka, Enrique Ruiz
Arriola and Mike Birse for many useful conversations.


\end{document}